\documentclass[sigconf]{acmart}
\usepackage{graphicx}
\usepackage{stfloats}
\usepackage{subfigure}
\usepackage{hyperref}
\usepackage{url}

\AtBeginDocument{%
  \providecommand\BibTeX{{%
    \normalfont B\kern-0.5em{\scshape i\kern-0.25em b}\kern-0.8em\TeX}}}

\setcopyright{acmcopyright}
\copyrightyear{2022}
\acmYear{2022}
\acmDOI{XXXXXXX.XXXXXXX}

\acmConference[CIKM ’22]{31st ACM International Conference on Information and Knowledge Management}{October 17-22, 2022}{Georgia, USA}
\acmPrice{15.00}
\acmISBN{978-1-4503-XXXX-X/18/06}

\begin{document}

\title{Hammer PDF: An Intelligent PDF Reader for Scientific Papers}

\author{Sheng-Fu Wang}
\authornote{Both authors contributed equally to this research.}
\author{Shu-Hang Liu}
\authornotemark[1]
\email{{wangsf, liush}@bit.edu.cn}
\affiliation{%
  \institution{Beijing Institute of Technology}
  \city{Haidian Distr.}
  \state{Beijing}
  \country{China}
}

\author{Tian-Yi Che}
\email{ccty@bit.edu.cn}
\affiliation{%
  \institution{Beijing Institute of Technology}
  \city{Haidian Distr.}
  \state{Beijing}
  \country{China}
}

\author{Yi-Fan Lu}
\email{luyifan@bit.edu.cn}
\affiliation{%
  \institution{Beijing Institute of Technology}
  \city{Haidian Distr.}
  \state{Beijing}
  \country{China}
}

\author{Song-Xiao Yang}
\email{yangsongxiao0616@gmail.com}
\affiliation{%
  \institution{Beijing Institute of Technology}
  \city{Haidian Distr.}
  \state{Beijing}
  \country{China}
}

\author{Heyan Huang}
\email{hhy63@bit.edu.cn}
\affiliation{%
  \institution{Beijing Institute of Technology}
  \city{Haidian Distr.}
  \state{Beijing}
  \country{China}
}

\author{Xian-Ling Mao}
\authornote{Corresponding author}
\email{maoxl@bit.edu.cn}
\affiliation{%
  \institution{Beijing Institute of Technology}
  \city{Haidian Distr.}
  \state{Beijing}
  \country{China}
}

\renewcommand{\shortauthors}{S.F. Wang and S.H. Liu, et al.}

\begin{abstract}

It is the most important way for researchers to acquire academic progress via reading scientific papers, most of which are in PDF format.
However, existing PDF Readers like Adobe Acrobat Reader and Foxit PDF Reader are usually only for reading by rendering PDF files as a whole, and do not consider the multi-granularity content understanding of a paper itself.
Specifically, taking a paper as a basic and separate unit, existing PDF Readers cannot access extended information about the paper, such as corresponding videos, blogs and codes.
Meanwhile, they cannot understand the academic content of a paper, such as terms, authors, and citations.
To solve these problems, we introduce Hammer PDF, an intelligent PDF Reader for scientific papers.
Apart from basic reading functions, Hammer PDF has the following four innovative features:
(1) information extraction ability, which can locate and mark spans like terms and other entities;
(2) information extension ability, which can present relevant academic content of a paper, such as citations, references, codes, videos, blogs, etc;
(3) built-in Hammer Scholar, an academic search engine based on academic information collected from major academic databases;
(4) built-in Q\&A bot, which can find helpful conference information;
The proposed Hammer PDF Reader can help researchers, especially those studying computer science, to improve the efficiency and experience of reading scientific papers.
We have released Hammer PDF, available at \url{https://pdf.hammerscholar.net/face}.


\end{abstract}



\maketitle

\section{Introduction}
\label{section:Introduction}
Nowadays, researchers have to spend considerable time on reading scientific papers to keep abreast of all the latest developments concerning their specialized fields. Most of the literature is in PDF format, and there are many tools available on the market that support reading PDF documents, such as Adobe Acrobat Reader\footnote{\url{https://www.adobe.com/acrobat/pdf-reader.html}} and Foxit PDF Reader\footnote{\url{https://www.foxit.com/pdf-reader/}}.

However, existing PDF Readers like Adobe Acrobat Reader and Foxit PDF Reader are usually only for reading by rendering PDF files as a whole, and do not consider the multi-granularity content understanding of a paper itself. Specifically, taking a paper as a basic and separate unit, existing PDF Readers cannot access extended information about the paper, such as corresponding videos, blogs and codes. Meanwhile, they cannot understand the academic content of a paper, such as terms, authors and citations. For example, when a paper is opened through Adobe Reader, researchers can only read the paper itself, and cannot obtain the extended content such as its corresponding videos, tutorial blogs and implementation codes. If a researcher wants to know: "Where is the corresponding code/video/blogs? What is the meaning of a word? What about authors? Where is the full text of a reference?", he has to use other tools such as browsers, translators, web search engines or scholar search engines to find the answers, which is tedious and tend to interrupt the reading process, and thus is very low efficient. Why do not we use only one tool to accomplish all these functions?

Thus, to tackle these above problems, this paper will introduce Hammer PDF, a multi-platform intelligent PDF Reader for scientific papers, to improve the efficiency and experience of reading PDF documents through machine learning and academic search.
The proposed Hammer PDF Reader is available for both web and desktop applications (Windows, macOS, and Linux).
Hammer PDF has four new features as follows:


\begin{itemize}

\item Information extraction ability, which can first get key spans like terms, authors, and citations by information extraction methods, and then mark these spans on the view panel, enabling users to interact directly with these spans.
\item Information extension ability, which can present related academic information of a paper such as authors, citations, references, codes, videos, and blogs.
\item Built-in Hammer Scholar, which is an one-stop academic search engine based on academic information collected from major academic databases.
\item Built-in Q\&A bot, which can find useful conference information, such as host place, host date, and impact factor.

\end{itemize}

\begin{figure*}[ht]

\centering
\includegraphics[width=\linewidth]{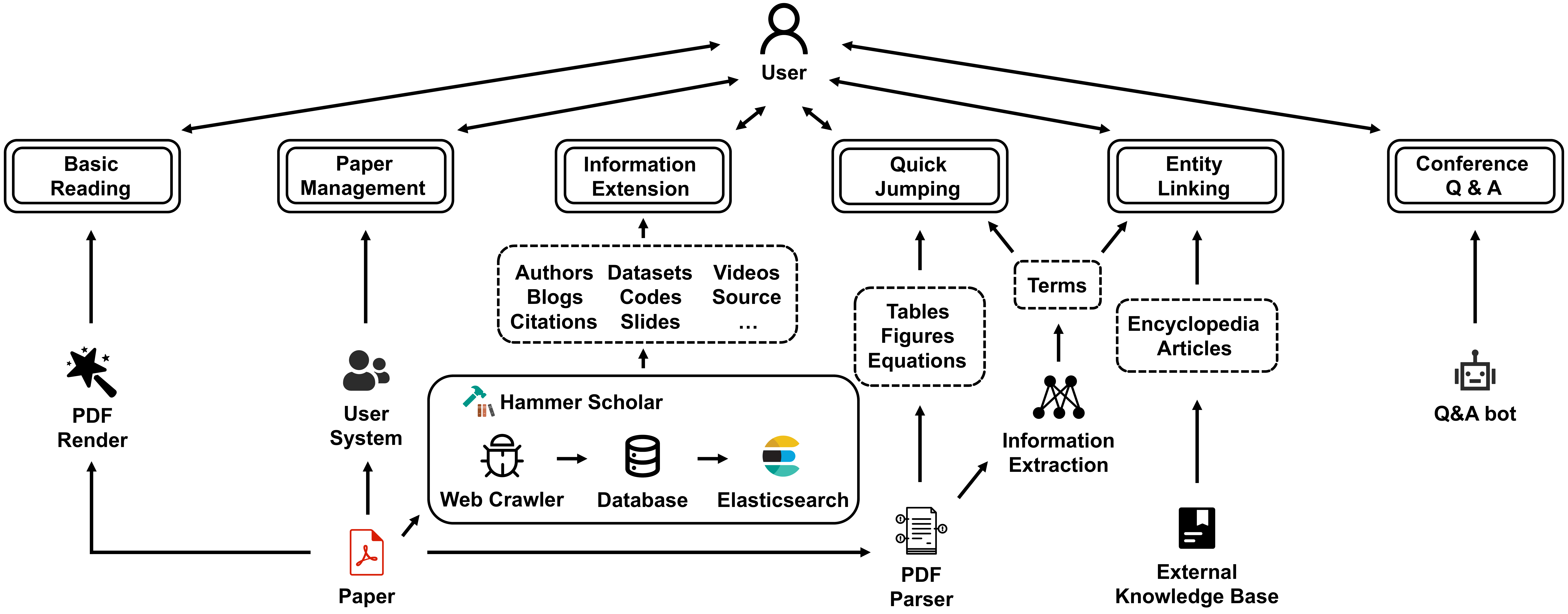}
\caption{The architecture and workflow of Hammer PDF.}
\label{fig_overview}

\end{figure*}

\section{Related Work}
\label{section:Related Work}
Traditional PDF Readers like Adobe Acrobat Reader and Foxit PDF Reader only support rendering, reading, and other basic functions. Meanwhile, these PDF Readers do not perform any content analysis, which fails to meet the intelligent needs of reading academic papers. As a result, several new PDF Readers focusing on mining academic value of scientific papers have come out recently.

ScholarPhi \cite{head2021augmenting, kang2020document} argues that it is often difficult to read articles when the information researchers need to understand them is scattered across multiple paragraphs. Thus, ScholarPhi\footnote{\url{https://www.semanticscholar.org/product/semantic-reader}} identifies terms and symbols in an article through fine-grained analysis and allows users to access their definitions. ReadPaper\footnote{\url{https://readpaper.com/}} focuses on academic communication through literature management and notes sharing, along with simple paper search functions through coarse-grained analysis. However, the academic database of ReadPaper is lacking in content to meet complex reading demands.

Compared to existing PDF Readers, Hammer PDF, introduced in this paper, is able to perform multi-granularity analysis of scientific papers, providing academic enhancements as intelligent features.

\section{Overview}
\label{section: Overview}
In this section, we introduce the overall architecture and core features of Hammer PDF.
As shown in Figure \ref{fig_overview}, Hammer PDF features basic reading as well as academic enhancements.
Specifically, academic enhancements include information extraction, information extension, academic search and conference Q\&A.
Section \ref{section:Intelligent Features} provides more details on academic enhancements.

\begin{figure*}[ht]

\centering
\includegraphics[width=\linewidth]{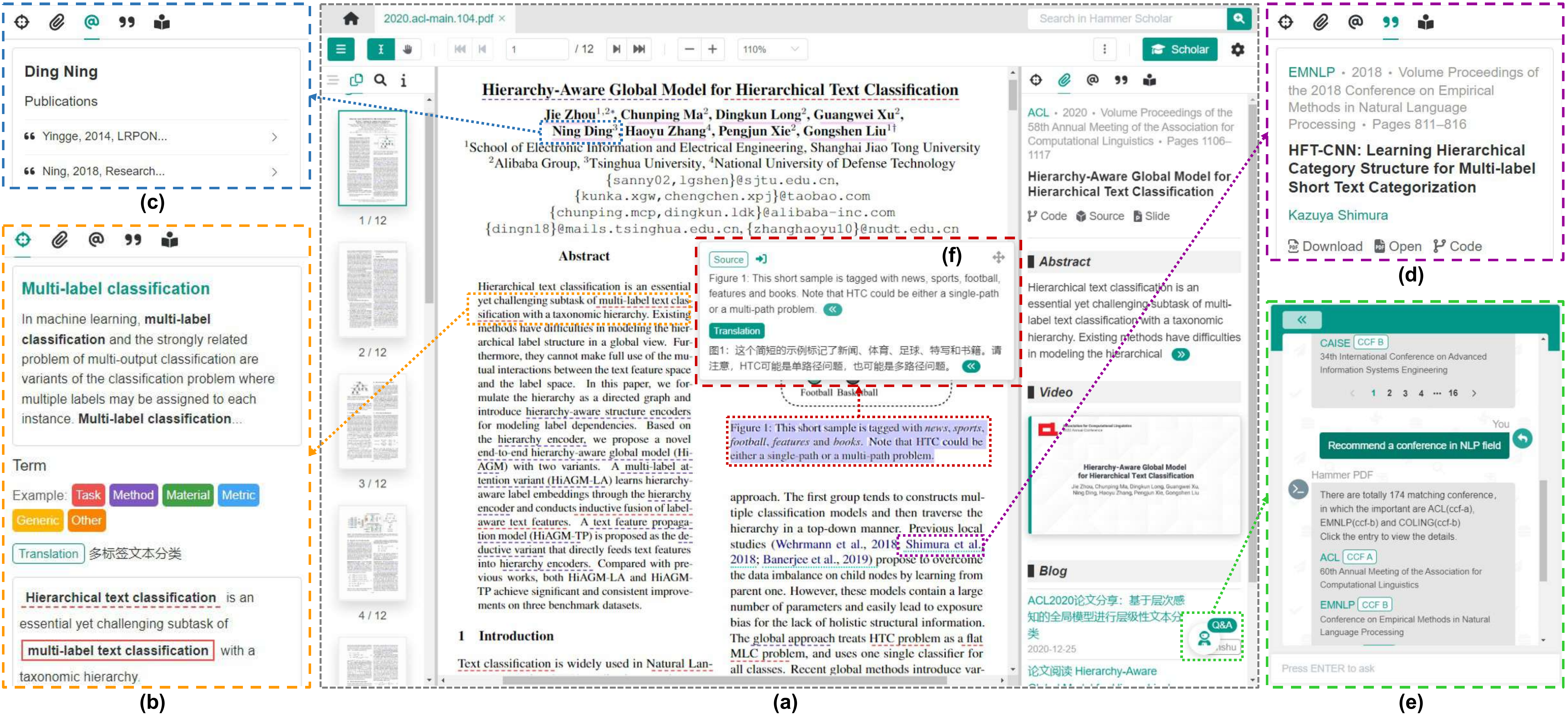}
\caption{The main function demonstration of Hammer PDF.}
\label{fig_user_case}

\end{figure*}

Hammer PDF offers basic reading functions.
Users can open PDF documents via local file, URL address, or DOI (Digital Object Identifier).
Also, users can open documents from the search results of the academic search engine Hammer Scholar\footnote{\url{https://hammerscholar.net/}}.
When a document is successfully opened, the interface will create a new tab called the view panel, as shown in Figure \ref{fig_user_case} (a).
The middle part of the view panel displays the document itself, while the Basic Sidebar on the left side provides text search and shows document information including outline, thumbnail, and metadata.
Also, on the right side of the view panel, the Academic Sidebar presents additional content for academic enhancements.
Moreover, users can access translations by directly selecting text within a document, which helps foreign researchers quickly understand the content.
As Figure \ref{fig_user_case} (f) shows, when some text is selected, a translation card is displayed near the text and can be dragged as needed.
In particular, users can change the translation service in the settings.

Hammer PDF also has simple document management capabilities, where users can open, bookmark, and delete one or more document records.
Furthermore, Hammer PDF is a multi-platform PDF Reader available for web and desktop applications.
In order to satisfy the reading needs of as many users as possible, the interface supports four languages, namely Simplified Chinese, Traditional Chinese, English and Japanese.

\section{Intelligent Features}
\label{section:Intelligent Features}

In this section, we introduce four features of Hammer PDF for academic enhancements, including information extraction, information extension, academic search and conference Q\&A.

\subsection{Information Extraction}
\label{subsection:Information Extraction}

To structure documents, we use Grobid \cite{lopez2009grobid, romary2015grobid} to get the logical structure of PDF documents.
Grobid is a machine learning library for extracting and parsing raw PDF files into structured documents, with an F1-score of 0.89 in parsing references \cite{tkaczyk2018machine}.
Then, we feed the extracted title, abstract and body into the information extraction model SpERT \cite{eberts2020span}, a span-based joint entity and relation extraction model.
Specifically, we use SpERT to perform NER (Named Entity Recognition) on the SciERC\cite{luan2018multi} dataset to obtain semantically rich terms and their types, including Task, Method, Metric, Material, Generic and Other.
SpERT achieves an F1-score of 70.33\% using SciBERT\cite{beltagy2019scibert} as a pre-trained language model for better results in scientific papers.




With these terms in place, we identify the page number and page location for each term.
Next, we mark an interactive underline mark where the term is located.
Figure \ref{fig_user_case} (b) depicts that different types of terms will be distinguished by different colored underlines.
The Academic Sidebar shows all terms in the document and their respective contexts.
When the user clicks on a term span, the sidebar displays the translation and other locations for the same term.

We can also capture author spans and citation spans from the structured document, and users are able to interact with these spans in the view panel just like term spans, as shown in Figure \ref{fig_user_case} (b).
When an author span is clicked, the Academic Sidebar presents the published works of the author.
When a citation span is clicked, the Academic Sidebar displays the details of the corresponding reference.

\subsection{Information Extension}
\label{subsection:Information Extension}

Based on the academic database provided by Hammer Scholar, which will be described in Section \ref{subsection:Built-in Academic Search}, we implement information extension with multiple academic features.
When the user opens a document, we retrieve the academic database for a matching paper according to the document's structure information.
Next, the Academic Sidebar presents the information of the retrieved paper, including the title, authors, abstract, citations, references, etc., as depicted in Figure \ref{fig_user_case} (a).
If the paper has related videos, tutorial blogs, or implementation codes attached, they will be presented as well.
Besides, users can click on the name of an author or the publication source to directly view its extended information.

We can perform information extension not only for papers but also for terms.
For example, when a term is selected, we fetch the relevant encyclopedia from Wikipedia\footnote{\url{https://www.wikipedia.org/}} and present it on the Academic Sidebar.
Specifically, for terms that cannot be matched perfectly on Wikipedia, we offer several partial matches as an alternative.
In addition, figures, tables, and equations in the paper support quick jumping, meaning that users can jump to the target by clicking the corresponding button in the Academic Sidebar.

\subsection{Built-in Academic Search}
\label{subsection:Built-in Academic Search}

We collect academic resources from six literature databases including arXiv\footnote{\url{https://arxiv.org/}}, ACL Anthology\footnote{\url{https://aclanthology.org/}} and DBLP\footnote{\url{https://dblp.org/}}.  
Academic resources contain the title, author, publication date, publisher, DOI, abstract, etc. 
Furthermore, we also collect presentations, blogs, videos, codes, and other extended resources for information extension. Over these resources, we build an academic search engine, named Hammer Scholar.

\begin{figure}[t]

\centering
\includegraphics[width=\linewidth]{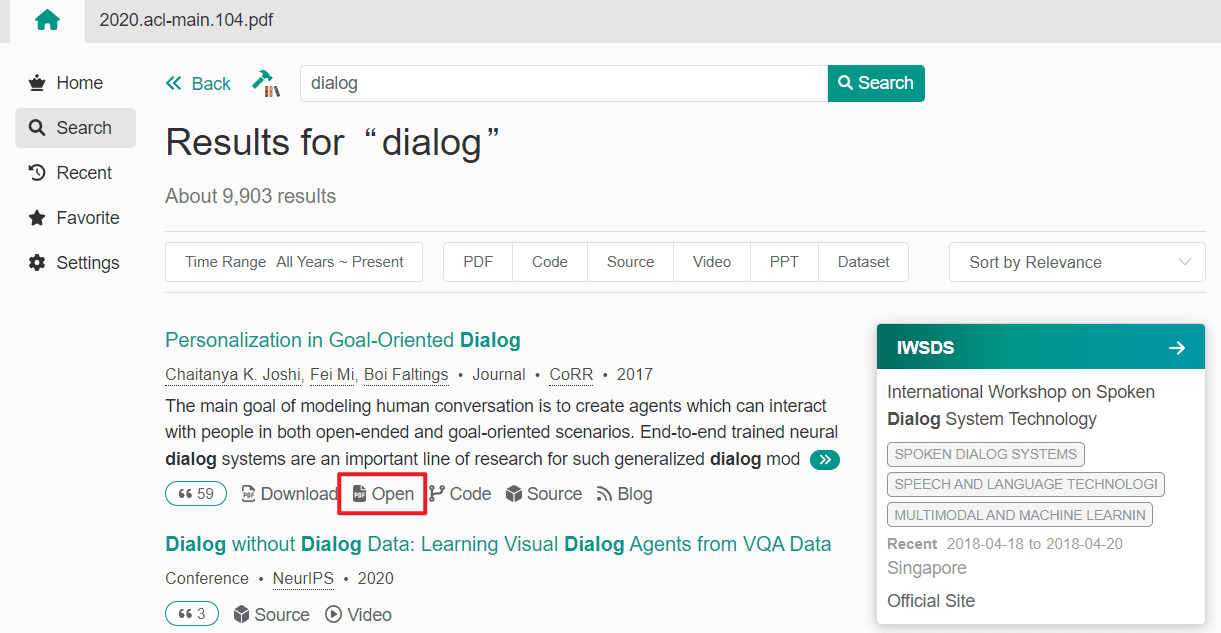}
\caption{Search results for papers in Hammer Scholar}
\label{fig_search_result}

\end{figure}

Hammer Scholar has a separate interface that provides both paper search and video search.
Take paper search as an example, the returned results after entering the keyword "dialog" are shown in Figure \ref{fig_search_result}.
Apart from filtering and sorting current search results, users can also pick a paper of interest and read it directly in the view panel by click "Open" button in a result, eliminating the need to find and upload the document.

\subsection{Built-in Q\&A Bot}
\label{subsection:Conference Q&A}

\begin{figure}[t]

\centering
\includegraphics[width=\linewidth]{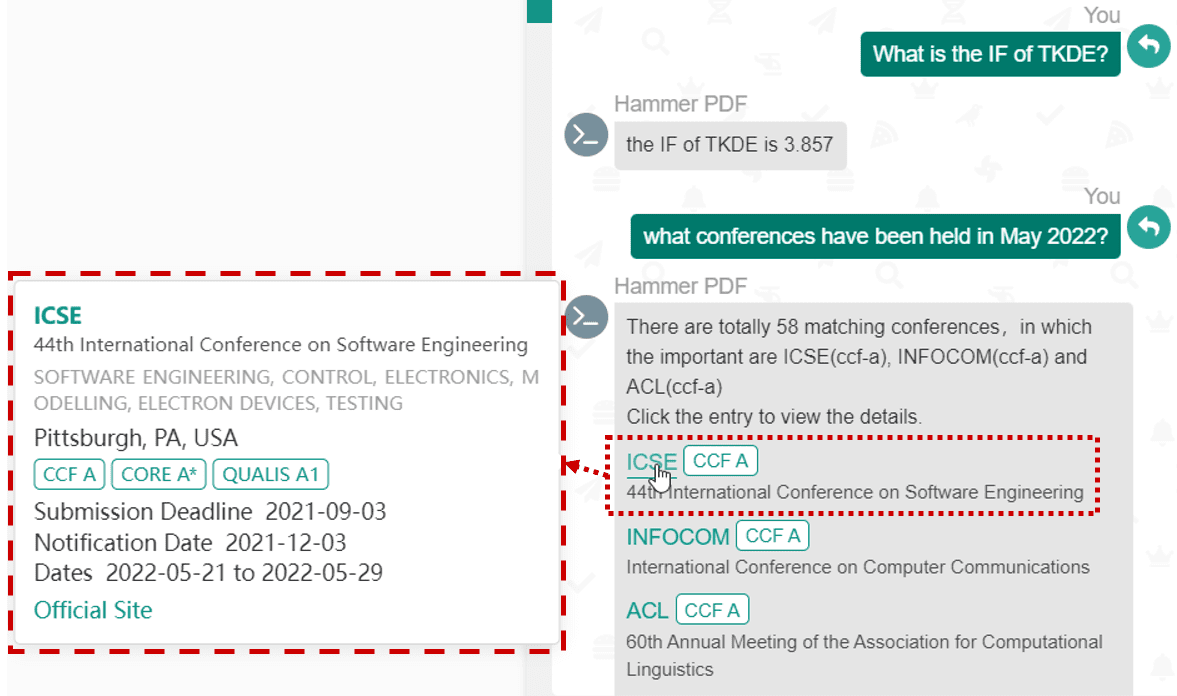}
\caption{Conference information query in academic Q\&A}
\label{fig_talk_panel}

\end{figure}

Academic Q\&A serves as a supplement to information extension and can answer questions related to conferences. 
We design several conference-related questions, such as host date, host place, deadline, conference level, impact factor, etc.
After the user asks a question, academic Q\&A identifies the question's intent.
If the question belongs to a conference recommendation, then the Q\&A model returns the answer in natural language.
If the question belongs to an academic search, the interface will jump to Hammer Scholar and search directly with the entered keywords.
As depicted in Figure \ref{fig_talk_panel}, when the user asks "What is the IF of TKDE?", the Q\&A model is able to reply the correct impact factor of the academic journal, while when asking "what conferences have been held in May 2022?", it returns a list of eligible conferences.
When the mouse hovers over the name of a conference, the user can view the conference's details.

To take previous conversations into account during the current conversation, we also design a multi-round conversation logic.
For instance, if a user asks “What is the deadline of ACL” and then asks “Where is it held”, the Q\&A model is able to correctly say where ACL is held.
Notably, academic Q\&A supports both Chinese and English to serve more users.


\section{Demonstration}
\label{section: Demonstration}

We demonstrate several core features of Hammer PDF through a comprehensive use case while using the figures from Section \ref{section:Intelligent Features} due to space constraints. First, users have 3 ways to open an academic article in PDF format: (1) upload a local file; (2) enter a URL address of a paper; (3) click "Open" button in returned results from built-in Hammer Scholar search engine, as shown in Figure \ref{fig_search_result}. 
Then, users can browse the document in the view panel and check the document information in the Basic Sidebar, as seen in Figure \ref{fig_user_case} (a). After that, the document is structured and parsed for academic enhancements. Once the document has been processed, the Academic Sidebar displays extended information about the paper, and users can switch the top navigation to view all terms within the article, as shown in Figure \ref{fig_user_case} (b). While reading, users can click on an author mark to see the author's detailed information (Figure \ref{fig_user_case} (c)), translate the selected text (Figure \ref{fig_user_case} (f)), or click on a citation mark to see the reference's details, as depicted in Figure \ref{fig_user_case} (d). Furthermore, users are also able to click on the "Open" button shown in Figure \ref{fig_user_case} (d) to access the full text of the reference of interest directly in a new view panel. 

Users who need to find a specific academic resource can use the built-in academic search engine to search for it. As illustrated in Figure \ref{fig_search_result}, for a certain search result, users can click on the title to view its details. Users can also chat with the conference Q\&A bot to obtain detail information about the academic journals or academic conference, as depicted in Figure \ref{fig_talk_panel}.


\section{Conclusion}
\label{section:Conclusion}

In this paper, we introduce Hammer PDF, a novel multi-platform intelligent PDF Reader for scientific papers.
Hammer PDF attempts to meet the growing intelligent needs of researchers during reading PDF documents. 
In addition to basic reading functions, we improve reading efficiency through information extraction, information extension, academic search, and conference Q\&A.
With lots of academic data, Hammer PDF can boost researchers' experience when reading scientific papers.

The introduction video and usage examples can be found on \url{https://pdf.hammerscholar.net/face}.
We recommend using browsers based on Chromium\footnote{\url{https://www.chromium.org/chromium-projects/}} to visit this site. 
Users can download desktop applications (Windows, macOS, and Linux) and submit feedback on \url{https://github.com/HammerPDF/Smart-Scientific-Reader}.


\bibliographystyle{ACM-Reference-Format}
\bibliography{main}

\end{document}